\documentclass{article}
\usepackage[T1]{fontenc} 
\usepackage[utf8]{inputenc} 
\usepackage{ismir,amsmath,cite,url}
\usepackage{graphicx}
\usepackage{color}

\usepackage{multirow}
\usepackage{amssymb}
\usepackage{enumitem}

\usepackage{xcolor}

\newcommand{\yhyangtext}[1]{{\color{black}#1}}

\title{A Benchmarking Initiative for Audio-domain Music Generation using the FreeSound Loop Dataset}

\multauthor
{Tun-Min Hung$^{1}$  \hspace{1cm} Bo-Yu Chen$^1$ \hspace{1cm} Yen-Tung Yeh$^{1, 2}$ \hspace{1cm} Yi-Hsuan Yang$^{1, 3}$} 
{\\
$^1$~Academia Sinica,  
$^2$~National Taiwan University,   
$^3$~Taiwan AI Labs \\
{\tt\small allenhung@iis.sinica.edu.tw, bernie40916@gmail.com, b06611042@ntu.edu.tw, yang@citi.sinica.edu.tw}
}

\def\authorname{TM Hung, BY Chen, YT Yeh, and YH Yang}

\usepackage[bookmarks=false,pdfauthor={\authorname},pdfsubject={\papersubject},hidelinks]{hyperref}

\begin{document}

\maketitle
\begin{abstract}
This paper proposes a new benchmark task for
generating musical passages in the audio domain by using the drum loops from the FreeSound Loop Dataset, which are publicly re-distributable.
Moreover, we use a larger collection of drum loops from Looperman to establish four model-based objective metrics for evaluation, releasing these metrics 
as a library for quantifying and facilitating the progress of musical audio generation. 
Under this evaluation framework, we benchmark the performance of three recent deep generative adversarial network (GAN) models we customize to generate loops, including StyleGAN, StyleGAN2, and UNAGAN. 
We also report a subjective evaluation of these models.
Our evaluation shows that the one based on StyleGAN2 performs the best in both objective and subjective metrics.
\end{abstract}

\section{Introduction}
\label{sec:introduction}



Audio-domain music generation
involves generating musical sounds either directly as audio waveforms 
or as time-frequency representations such as the Mel spectrograms.
Besides modeling musical content in aspects such as pitch and rhythm, it has the additional complexity of modeling the spectral-temporal properties of musical sounds, compared to its symbolic-domain music generation counterpart. 
In recent years, deep learning models have been proposed for audio-domain music generation, starting with simpler tasks such as generating  instrumental single notes \cite{engel2017neural,engel2019gansynth,drumgan,9053128}, a task also known as \emph{neural audio synthesis}. 
Researchers have also begun to address the more challenging setting of generating sounds of longer duration \cite{oord16,samplernn,dadabots,melnet,liu2020unconditional,liu2019score,dhariwal2020jukebox,mp3net}. 
For example, Jukebox \cite{dhariwal2020jukebox} aims to generate realistic minutes-long singing voices conditioned on lyrics, genre, and artists; and UNAGAN \cite{liu2020unconditional} aims to generate musical passages of finite yet arbitrary duration for singing voices, violin, and piano, in an unconditional fashion.

\begin{figure}[t]
\centering
\includegraphics[width=.84\columnwidth]{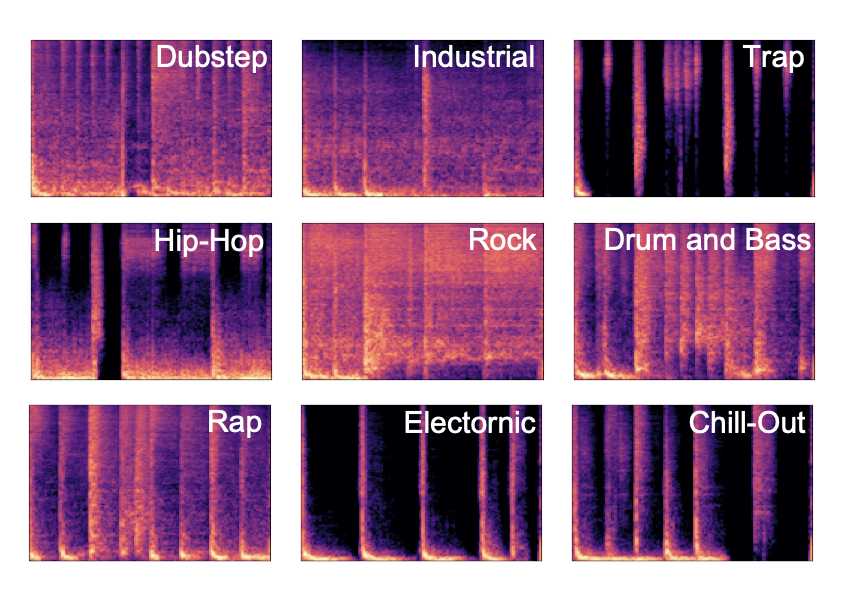}
\caption{The mel-spectrograms of some random drum loops generated by the StyleGAN2 model \cite{9156570} trained on the looperman dataset, with the genre labels predicted by the short-chunk CNN \cite{won2020evaluation} classifier (see Section \ref{sec:metrics:IS}). }
\label{fig:melspec}
\end{figure}

The focus of this paper is on the evaluation  of audio-domain music generation. We note that, for model training and evaluation, research on generating single notes quite often adopts NSynth \cite{engel2017neural}, a large public dataset consisting of individual notes from different instruments. 
The use of a common dataset for evaluation ensures the validity of performance comparison between different models. Such a standardized dataset for benchmarking, however, is not available when it comes to generating longer musical passages, to our best knowledge.
Oftentimes private in-house datasets are employed in existing works; for example, both UNAGAN \cite{liu2020unconditional} and Jukebox \cite{dhariwal2020jukebox} employ audio recordings scrapped from the Internet, which cannot be shared publicly. 
The only exception is MAESTRO, a public dataset with over 172 hours of solo piano performances, employed by MelNet \cite{melnet}, UNAGAN \cite{liu2020unconditional}, and MP3net \cite{mp3net}.
However, MAESTRO is piano-only so not diverse enough in  timbre.



We see new opportunities to address this gap with the recent release of the FreeSound Loop Dataset (FSLD) \cite{ramires2020freesound}, which contains 9,455 production-ready, public-domain loops distributed under Creative Commons licenses.\footnote{\url{https://zenodo.org/record/3967852}} 
We therefore propose to use \emph{audio-domain loop generation}, a task seldom reported in the literature, to set a benchmark for musical audio generation research.

We deem loops as an adequate target for audio generation for their following merits.
First, loops are audio excerpts, usually of short duration, that can be played in seamless manner\cite{ramires2020freesound,LoopsasGenreResources}. Hence, the generated loops can be played repeatedly. 
Second, loops are fundamental units in the production of many contemporary dance music genres.
A loop can usually be associated with a single genre or instrument label \cite{ramires2020freesound}, and a certain ``role'' (e.g., percussion, FX, melody, bass, chord, voice) \cite{joannching}.
Third, loops are fairly diverse in their music content and timbre, as sound design has been a central part in making loops.


A primary contribution of this paper is therefore the proposal and implementation of using FSLD as a benchmark for audio generation. In particular, we adapt three recent deep generative adversarial network (GAN) \cite{GAN} models and train them on the \emph{drum-loop} subset of FSLD,
and report thorough evaluation of their performance, both objectively and subjectively.
This includes UNAGAN \cite{liu2020unconditional} and two state-of-the-art models for image generation, StyleGAN \cite{8953766} and StyleGAN2\cite{9156570}.

Drum loop generation is interesting in its own right due to its applications in automatic creation of loop-based music \cite{paul20ismir}. 
As \cite{richard17} indicates, drum beats represent one of the most critical and fundamental elements that form the style of EDM. Moreover, drum loops are already fairly diverse in musical content,  
as demonstrated in Figure \ref{fig:melspec}.
Although we only consider drum loops here for the sake of simplicity, this benchmark can be easily extended to cover all the loops from FSLD in the near future.





Our secondary contribution lies in the development of standardized objective metrics for evaluating audio-domain loop generation, which can be equally important as having a standardized dataset.
We collect a larger drum loop dataset from an online library called looperman,\footnote{\url{https://www.looperman.com/}} with roughly 9 times more drum loops than FSLD, and use this looperman dataset to build four model-based  metrics (e.g., inception score \cite{DBLP:conf/nips/SalimansGZCRCC16})\footnote{We refer to them as \emph{model-based} metrics because we need to build a classifier or a clustering model to calculate the metrics; see Section \ref{sec:metrics}.} to evaluate the acoustic \emph{quality} and \emph{diversity} of the  loops generated by 
the GAN models.
While this looperman dataset cannot be released publicly due to copyright concerns,
we release the metrics and the trained GAN models for drum loop generation at the following GitHub repo:  \url{https://github.com/allenhung1025/LoopTest}.

Moreover, we put some of the generated drum loops on an accompanying  \textbf{demo website},\footnote{\url{https://loopgen.github.io/}}
which we recommend readers to visit and listen to.
We also present the result where we use the method of \emph{style-mixing} of StyleGAN2 to generate ``interpolated'' versions of loops.

Below, we review related work in Section \ref{sec:related}, present the datasets in Section \ref{sec:db}, the proposed objective metrics in Section \ref{sec:metrics}, the benchmarked  models in Section \ref{sec:model}, and  the evaluation result in Section \ref{sec:eval}.

\section{Related Work
}\label{sec:related}

Existing work on audio-domain music generation can be categorized in many ways. 
First, an \textbf{unconditional} audio generation model takes as input a vector $\mathbf{z} \in \mathbb{R}^{N_z}$ of a fixed number of random variables (or a sequence of such vectors; see below) and generates an audio piece from scratch.
When side information of the target audio to be generated is available, we can feed such prior information as another vector $\mathbf{c} \in \mathbb{R}^{N_c}$ and use it as an additional input to the generative model, making it a \textbf{conditional} generation model. For example, GANSynth \cite{engel2019gansynth} uses the pitch of the target audio as a condition. While we focus on unconditional generation in our benchmarking experiments presented in Section \ref{sec:eval}, it is straightforward to extend all the models presented in Section \ref{sec:model} to take additional conditions.

Second, some existing models can only generate \textbf{fixed-length} output, while others can do \textbf{variable-length} generation.
One approach to realize variable-length generation is by using as input to the generative model  a sequence of latent vectors $\mathbf{z}_1, \mathbf{z}_2, \dots$, instead of just one latent vector $\mathbf{z}$. This is the approach taken by UNAGAN \cite{liu2020unconditional}, Jukebox \cite{dhariwal2020jukebox}, and VQCPC-GAN \cite{nistal2021vqcpcgan}.

Third, existing models for generating single notes are typically \textbf{non-autoregressive} models \cite{engel2019gansynth,drumgan,9053128}, i.e., the target is generated at one shot. 
When it comes to generating longer phrases, \textbf{autoregressive} models, that generate the target piece one frame or one time sample at a time in the chronological order,
might perform better \cite{oord16,dhariwal2020jukebox}, as the output of such models depends explicitly on the previous frames (or samples) that have been generated.

Existing models have been trained and evaluated to generate different types of musical audio, including 
singing voice\cite{liu2020unconditional, liu2019score, dhariwal2020jukebox}, drum \cite{drumgan, 9053128, aouameur2019neural, drysdale2020adversarial}, violin \cite{liu2020unconditional}, and piano\cite{oord16, melnet,liu2020unconditional,mp3net}. 
The only work addressing loop generation is the very recent LoopNet model from Chandna \emph{et al.} \cite{loopnet}. They also use loops from looperman but not anything from FSLD or other public datasets, hence not constituting a benchmark for audio generation.

For drum generation in particular, work has been done in the symbolic domain to generate drum patterns \cite{DBLP:conf/icml/GillickREEB19,drumVAE,alain2020deepdrummer,tokui2020gan} and a drum track as part of a symbolic multi-track composition \cite{musegan,simon2018learning,ren2020popmag}. 
For example, 
DeepDrummer \cite{alain2020deepdrummer} employs human-in-the-loop to produce drum patterns preferred by a user. 
In the audio domain,
DrumGAN \cite{drumgan} and the model proposed by Ramires \emph{et al.} \cite{9053128} both work on only single hits, i.e., one-shot drum sounds.
They both use the Audio Commons models \cite{audiocommons}
to extract high-level timbral features to condition the generation process.
DrumNet \cite{lattner2019highlevel} is a model that generates
a sequence of (monophonic) kick drum hits, not the sounds of an entire drum kit.



\section{Datasets}\label{sec:db}



Two datasets are employed in this work. The first one is a subset of drum loops from the public dataset FSLD \cite{ramires2020freesound}, which is used to train the generative models for benchmarking. 
FSLD comes with detailed manual labeling of the loops with tags such as instrumentation, rhythm, tone and genre. 
As stated in the FSLD paper \cite{ramires2020freesound}, FSLD is balanced in terms of musical genre.
By picking loops which are tagged with the keywords ``drum'', ``drums'' or ``drum-loop'', we are able to find 2,608 drum loops out of the 9,455 loops available in FSLD. 
We do not need to hold out any of them as test data but use all these bars for training our generative models, since we focus on unconditional generation in this paper; i.e., each generative model will generate a set of loops randomly for evaluation.

\begin{figure}[t]
\centering
\includegraphics[trim={0 2.8mm 0 0},clip,width=0.95\columnwidth]{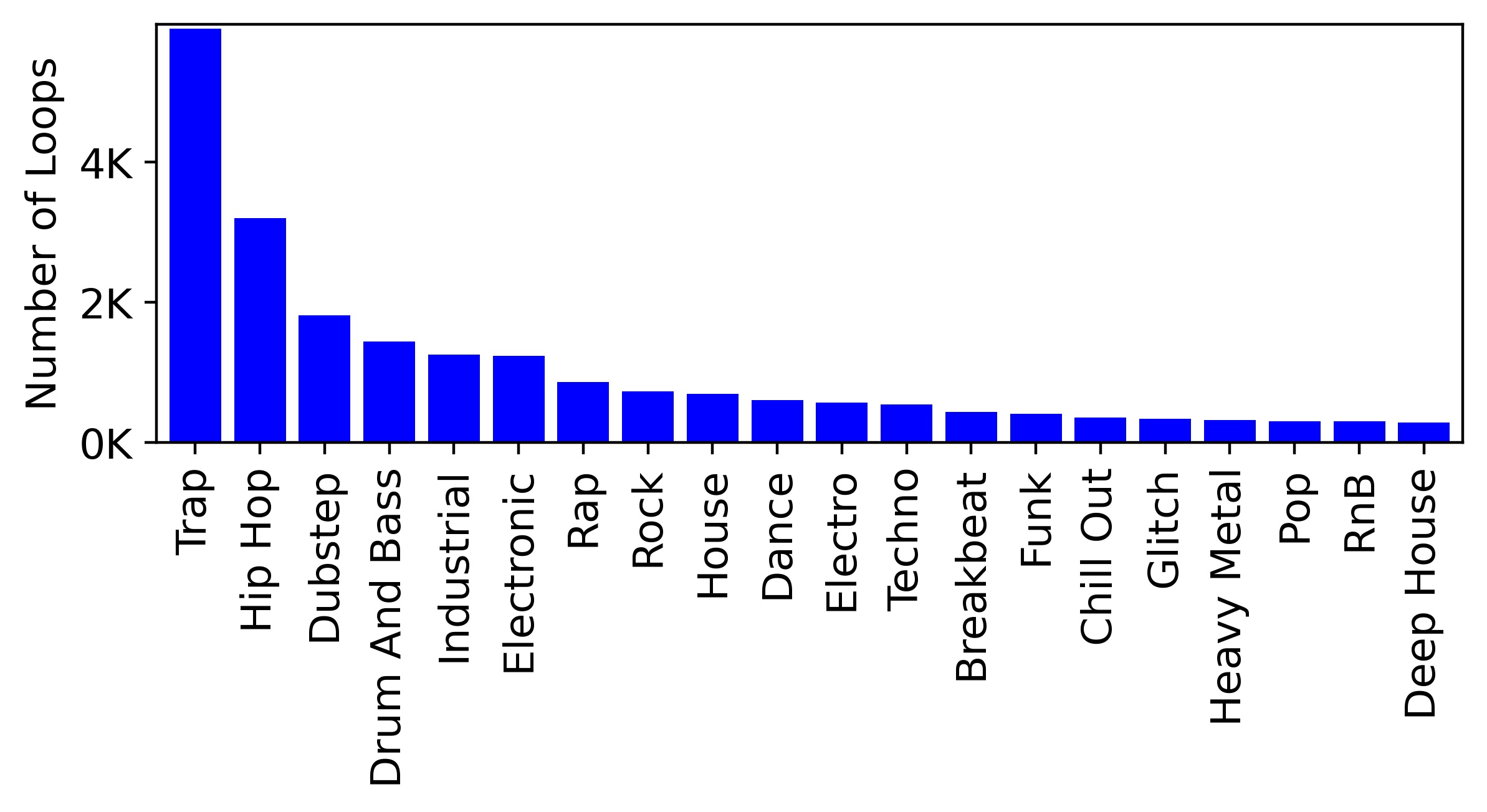}
\caption{Genre distribution of the drum loops from looperman; we display only the top 20 out of 66 genres.}
\label{fig:genre_loop}
\end{figure}

The second dataset is a larger, private collection of drum loops we collect from looperman, a website hosting free music loops.\footnote{As stated on \url{https://www.looperman.com/help/terms}, ``All samples and loops are free to use in commercial and non commercial projects.'' But, ``You may NOT use or re-distribute any media from the loops section of looperman.com as is either for free or commercially on any other web site.'' (Accessed August 1, 2021)} 
We are able to collect in total  23,983 drum loops, which is much more than the drum loops in FSLD. 
We use the looperman dataset mainly for establishing the model-based objective metrics for evaluation (see Section \ref{sec:metrics}). For instance, we train an audio-based genre classifier using looperman to set up the drum-loop version of the ``inception score''  \cite{DBLP:conf/nips/SalimansGZCRCC16,barratt2018note} to measure how likely a machine-generated loop sounds like a drum loop. 
Figure \ref{fig:genre_loop} shows the number of tracks per genre tag in looperman, which exhibits a typical long-tail distribution. We can see that ``Trap'' is the most frequent genre, with 5,903 loops.

We use looperman instead of FSLD to set up such objective metrics, since a larger dataset increases the validity and generalizability of the metrics.
Moreover, although we cannot re-distribute the loops from looperman according to its terms, we can share  checkpoints of the pre-trained models 
for computing the proposed objective metrics.


\subsection{Data Pre-processing}
As we are interested in benchmarking the performance of one-bar loop generation, we perform downbeat tracking using the state-of-the-art recurrent neural network (RNN) model available in the Madmom library \cite{10.1145/2964284.2973795,Bock2016d} to slice every audio file into multiple one-bar loops.\footnote{The downbeat tracker in Madmom is fairly accurate for percussive  audio such as the drum loops. For example, it reaches F1-score of 0.863 on the Ballroom dataset \cite{ballroom}, according to \cite{Bock2016d}.} 
After this processing, we have in total 13,666 and 128,122 one-bar samples from FSLD and looperman, respectively. 
We refer to these two collections of one-bar drum loops as the \textbf{freesound} and \textbf{looperman} datasets hereafter.
We note that all these one-bar samples are of four beats. 

\begin{figure}[t]
\centering
\includegraphics[trim={0 3mm 0 0},clip,width=0.95\columnwidth]{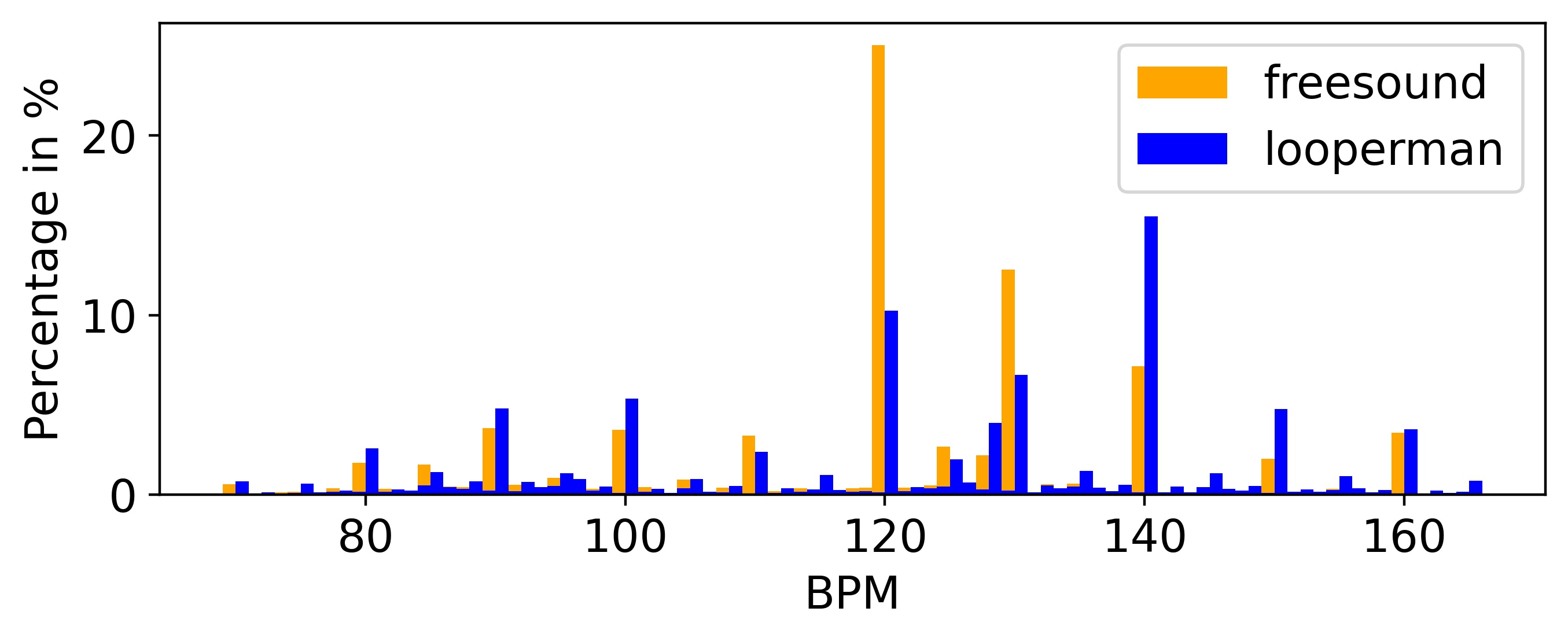}
\caption{Tempo distribution of the two sets of loops. Y axis represents the percentage of all loops in the dataset.}
\label{fig:bpm}
\end{figure}

As shown in Figure \ref{fig:bpm}, the one-bar samples in either the freesound or looperman datasets have different tempos and hence different lengths.
To unify their length to facilitate benchmarking, we use pyrubberband\footnote{\url{https://pypi.org/project/pyrubberband/}} to temporally stretch each of them to 2-second long, namely to have 120 BPM (beat-per-minute) as their tempo. We listened to some of the stretched samples in both datasets and found most sounded plausible with little perceptible artifacts.\footnote{This, however, may not be the case if the loops are not drum loops. Some data filtering might be needed then, e.g., to remove those whose tempo are much away from 120 BPM.}

All the loops are in 44,100 Hz sampling rate. We down-mix the stereo ones into mono. 
After that, we follow the setting of UNAGAN \cite{liu2020unconditional} 
to compute the Mel spectrograms of these samples, with 
1,024-point window size hann window and 275-point hop size for short-time Fourier Transform (STFT), and 80 Mel channels.




\section{Evaluation Metrics}
\label{sec:metrics}

We consider four metrics in our benchmark, developing the drum-loop version of them using the looperman dataset. 

\subsection{Inception Score (IS)}
\label{sec:metrics:IS}

IS \cite{DBLP:conf/nips/SalimansGZCRCC16,barratt2018note} measures the quality of the generated data and detects whether there is a mode collapse by using a pre-trained domain-specific classifier.
It is computed as the KL divergence between the conditional probability $p(y|x)$ and marginal probability $p(y)$, 
\begin{equation}
    \mathrm{IS}=\exp\big(E_x[ \text{KL} ( p(y|x) \| p(y)] \big) \,,
\end{equation}
where $x \in \mathcal{X}$ denotes a data example (e.g., a generated loop), and $y \in \mathcal{Y}$ is a pre-defined class. 
Specifically, the calculation of IS involves building a classifier over the type of data of interest, and it achieves high score (namely, the \emph{higher} the better) score when 1) each of the generated data can be classified to any of the predefined classes with high confidence, and 2) the generated data as a whole has close to uniform distribution over the predefined classes. 

We use looperman to establish such a classifier, using its genre labels for training a 66-class classifier over the Mel spectrograms of one-bar samples. Specifically, we split the data by 100,000/10,000/18,111 as the training, validation, and test sets, and use the state-of-the-art music auto-tagging model \emph{short-chunk CNN} \cite{won2020evaluation}\footnote{\url{github.com/minzwon/sota-music-tagging-models}} for model training. 
The classifier achieves 0.748 accuracy on the test set. 

\subsection{Fr\'{e}chet Audio Distance (FAD)}

The idea of FAD, as proposed by Kilgour \emph{et al.} \cite{kilgour2019frechet}, is to measure the closeness of the data distribution of the real data versus that of the generated data, in a certain embedding space. 
Specifically, they pre-train a VGGish-based audio classifier on a large
collection of YouTube videos for classifying 300$+$ audio classes and sound events, and then use the second last 128-dimension layer (i.e., prior to the final classification layer) for this embedding space \cite{kilgour2019frechet}.
The data distributions of $r$eal and $g$enerated data in this space are  modeled as a multi-variate normal distribution characterized by $(\mathbf{\mu}_r, \mathbf{\Sigma}_r)$ and $(\mathbf{\mu}_g, \mathbf{\Sigma}_g)$ respectively. The FAD score is then computed by the following equation,
\begin{equation}
    \mathrm{FAD} = \Vert \mathbf{\mu}_r - \mathbf{\mu}_g \Vert ^ 2 + tr(\mathbf{\Sigma}_r + \mathbf{\Sigma}_g - 2 \sqrt{\mathbf{\Sigma}_r \mathbf{\Sigma}_g}) \,,
\end{equation}
and is the \emph{lower} the better (down to zero).
We use the open source code and pre-trained classifier \footnote{\url{github.com/google-research/google-research/tree/master/frechet\_audio\_distance}} to compute the FAD, using the  looperman data as the real data 
and the  output of a generative model as the generated data.

\subsection{Diversity Measurement}
Following \cite{liu2020unconditional}, we measure diversity with the number of  statistically-different bins (NDB) and Jensen-Shannon divergence (JSD) metrics proposed by Richardson \emph{et al.} \cite{DBLP:conf/nips/RichardsonW18}, via the official open source code.\footnote{\url{github.com/eitanrich/gans-n-gmms}}
We firstly run $K$-means clustering over normalized Mel spectrograms of 
10 thousands one-bar samples randomly picked from looperman
to get $K = 100$ clusters, and count the number of samples per cluster, $n_k$, for each $k$.
Then, given a collection of  loops randomly generated by a generative model, we fit the loops into the clustering and also count the number of fitted samples per cluster, $\widehat{n_k}$.
We can then measure the difference between the two distributions $\{n_k\}$ and $\{\widehat{n_k}\}$ by either the number of statistically-different bins (among the $K$ bins; the \emph{lower} the better) and their JSD (the \emph{lower} the better; down to zero).
Richardson \emph{et al.} \cite{DBLP:conf/nips/RichardsonW18} recommend reporting the value of NDB divided by $K$, saying that if the two samples do come from the same distribution, NDB$/K$ should be equal to the significance level of the statistical test, which we set to 0.05.


\begin{figure*}[t]
\centering
\includegraphics[width=0.88\textwidth]{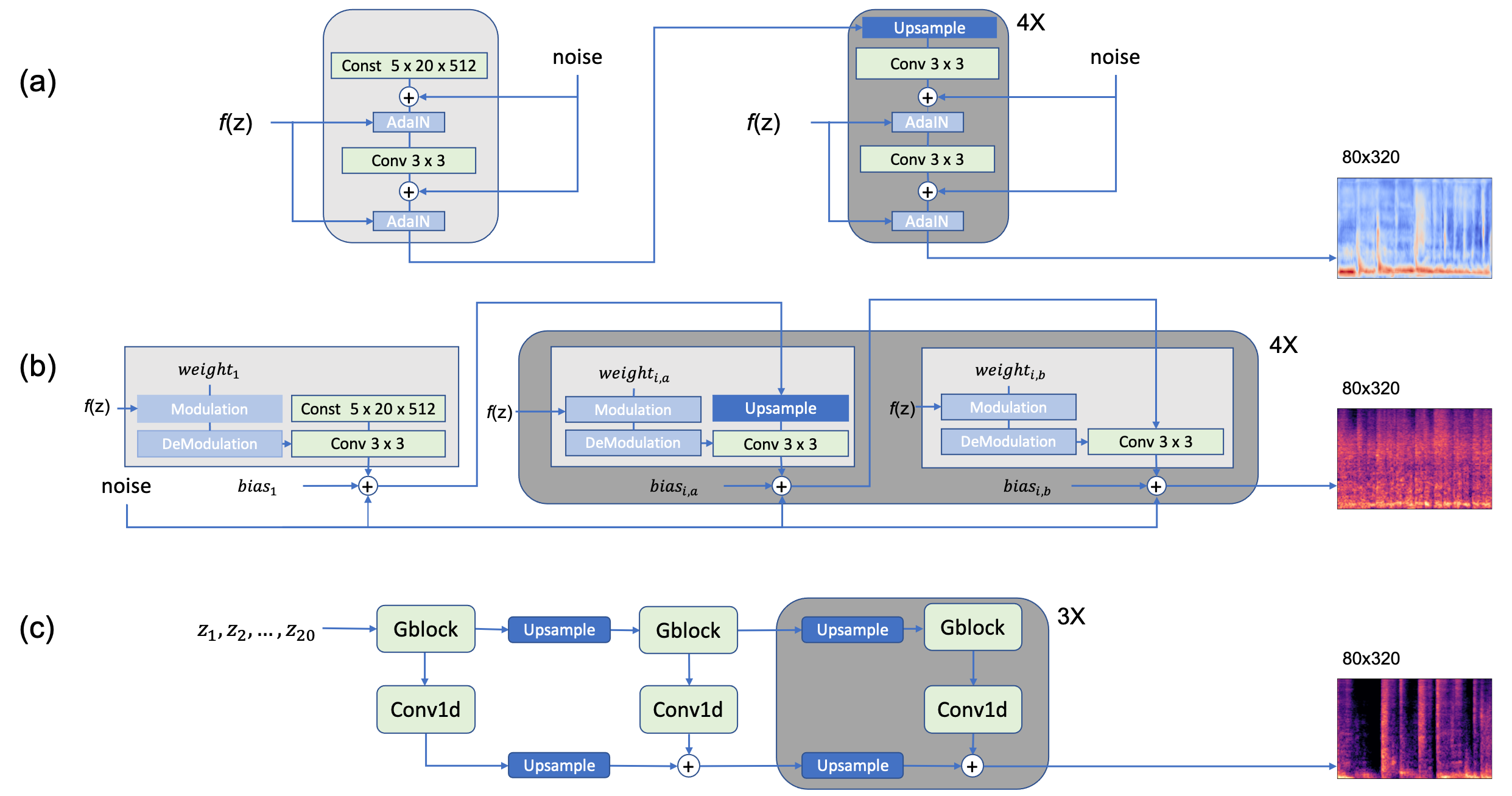}
\caption{Schematic plots of the adapted (a) StyleGAN~\cite{8953766}, (b) StyleGAN2~\cite{karras2018progressive}, and (c) UNAGAN~\cite{liu2020unconditional} in our benchmark. \yhyangtext{Only a single latent vector $\mathbf{z}$ is used as the input for the fully convolutional models in (a) and (b), while a sequence of 20 latent vectors are used in model (c), which uses a stack of gate recurrent unit (GRU) layer and grouped convolution layer in each of its `Gblocks'
\cite{liu2019score,liu2020unconditional}. In (b), $\text{weight}_i$ and $\text{bias}_i$ are parameters of the $3\times3$ convolution layers to be learned.}}
\label{fig:model}
\end{figure*}

\section{Benchmarked Generative Models}
\label{sec:model}

We develop and evaluate in total three recent deep generative models, 
all of which happen to be GAN-based \cite{GAN}.
The first model is \textbf{StyleGAN2} \cite{9156570}, which represents the state-of-the-art in image generation, included here intending to test its applicability for musical audio generation (which has not been reported elsewhere, to our best knowledge).
The second model, \textbf{StyleGAN} \cite{8953766}, is a precursor of StyleGAN2, tested on spoken digit generation before (akin to single note generation in music) \cite{palkama20interspeech} but not on musical audio generation. 
Both StyleGAN and StyleGAN2 generate only fixed-length output, which is fine here since our samples have constant length.
The last model, \textbf{UNAGAN} \cite{liu2020unconditional}, represents a state-of-the-art in musical audio generation, capable of generating variable-length output.
For fair comparison, we only require UNAGAN to generate two-second samples as the other two.
Schematic plots of the three models can be found in Figure \ref{fig:model}.

All these three models are trained to generate Mel spectrograms, with phase information missing. But, the Mel spectrograms can later be converted into audio waveforms by a separate \emph{neural vocoder}, such as WaveNet \cite{oord16}, WaveGlow \cite{prenger2018waveglow}, DiffWave \cite{kong2021diffwave}, or MelGAN \cite{melgan}.  
We are in favor of MelGAN for it is non-autoregressive and therefore fast in inference time, and for there is official open source code that is easy to use.\footnote{\url{github.com/descriptinc/melgan-neurips}}
We train MelGAN on the looperman dataset and use it in all our experiments.


\subsection{StyleGAN}
StyleGAN and StyleGAN2 are both non-autoregressive models for generating images. 
They take a constant tensor of size $4\times4\times512$ as input, and use a mapping network $f(\cdot)$ consisting of eight linear layers to map a random latent vector $\mathbf{z}$ to an intermediate lantent vector $\mathbf{w}$, which affects the generation process 
by means of adaptive instance normalization (AdaIN) operations in every block of the generator \cite{8953766}. 
Each bock progressively upsamples its input to a larger tensor, until reaching the target size of  $1024\times1024$ by the end with in total eight such blocks.

The input tensor of StyleGAN and StyleGAN2 can be interpreted as 512 $4\times4$ tiny images. This tensor is learned and then fixed during the inference stage while generating new images, using different $\mathbf{z}$ each time. 
We modify it to be a $5\times20\times512$ tensor in our work, to generate a $80\times320$ Mel spectrogram through four upsampling blocks.

Our implementation of StyleGAN is based on an open source code.\footnote{\url{github.com/rosinality/style-based-gan-pytorch}} For model training, StyleGAN employs 
the non-saturating loss with $R_1$ regularization\cite{mescheder2018training} and a progressive-growing training strategy \cite{karras2018progressive}.
We use 0.9 mixing regularization ratio \cite{karras2018progressive}, and set the batch size to 32, 16, 8, 4  in the respective scale, from low to high resolution. In every scale, we train with 1.2M samples. We deployed Adam optimization algorithm and set the learning rate to 1e--3. The total training time is 120 hr on an NVIDIA GTX1080 GPU with 8GB memory.

\subsection{StyleGAN2}
StyleGAN2 \cite{9156570} is an improved version of StyleGAN with many structural changes, 
\yhyangtext{including replacing AdaIN by a combination of ``modulation'' and ``demodulation'' layers,}
processing the input tensor differently, adding the Gaussian noise outside of the style blocks etc. 
\yhyangtext{The weights in the $3\times3$ convolution layers are scaled with $f(\mathbf{z})$ in the  Modulation block and normalized by L2 norm in the DeModulation block.} We refer readers to the original paper \cite{9156570} for details.
Our implementation of StyleGAN2 is based on another open source code,\footnote{\url{github.com/rosinality/stylegan2-pytorch}}  with similar training strategies as the StyleGAN case, but two times larger learning rate, no progressive growing, and a constant batch size of 8 for 1M samples. The total training time is 100 hr on a GTX1080.

\subsection{UNAGAN}
UNAGAN \cite{liu2020unconditional} is a non-autoregressive model originally designed for generating variable-length singing voices in an unconditional fashion.
The authors also demonstrate its effectiveness in learning to generate passages of violin, piano, and speech. 
What makes UNAGAN different from existing models such as StyleGAN, WaveGAN \cite{donahue19iclr}, DrumGAN \cite{drumgan}, and GANSynth \cite{engel2019gansynth} is that UNGAN takes a sequence of latent vectors $\mathbf{z}_1, \mathbf{z}_2, \dots$ as input, instead of just a single one. 
This sequence of latent vectors, together with the recurrent units inside its `Gblocks' \cite{liu2019score,liu2020unconditional} (see Figure \ref{fig:model}(c)), facilitates UNAGAN to generate variable-length audio with length proportional to the length of the input latent sequence. 
UNAGAN adopts a hierarchical architecture that generates Mel spectrograms in a coarse-to-fine fashion similar to the progressive upsampling blocks in StyleGAN  and StyleGAN2. 
UNAGAN uses 
the BEGAN-based adversarial loss \cite{berthelot17}, and an additional cycle consistency loss \cite{cyclegan} to stabilize training and for increasing diversity. 
Our implementation of UNAGAN is based on the official open source code.\footnote{\url{https://github.com/ciaua/unagan}}
We fix the number of input latent vectors to 20 and train the model with Adam, 1e--4 learning rate, and a batch size of 16 for 100k iterations, amounting to 40 hr on a GTX1080.

\vspace{-2mm}
\section{Evaluation}
\label{sec:eval}

\subsection{Objective Evaluation Result}
Table \ref{tab:exp-obj-freesound} presents the objective evaluation result of models trained on the \emph{freesound} dataset.
Each model generates 
2,000 random loops to compute the scores. 
We also compute these metrics on the two real datasets and add the results to Table \ref{tab:exp-obj-freesound}, to offer an oracle reference. We see that the IS of StyleGAN2 
is the closest to that of the freesound dataset, 
followed by UNAGAN and then StyleGAN. 
Student's $t$-test shows that the performance edge of StyleGAN2 over either UNAGAN or StyleGAN is statistically significant ($p$-value$<$0.01).
This reveals the efficacy of StyleGAN2 for generating fixed-length audio.


The scores in JS and NDB further support the superiority of StyleGAN2, showing that its output is the most diverse among the three. 

The scores in FAD, however, shows that UNAGAN performs better than StyelGAN2 here. The contrast between IS and FAD suggests that UNAGAN learns to generate samples whose embeddings have similar distribution as the real data, but its output cannot be easily associated with a genre class by the short-chunk CNN classifier.
We also see that StyleGAN has fairly high FAD, showing that its generation hardly resemble the real data distribution. 

Out of curiosity, we also train the  models on the private, yet larger, \emph{looperman} dataset and redo the evaluation. Table \ref{tab:exp-obj-looperman} shows that StyleGAN2 achieves even higher IS and much lower NDB here. Furthermore, its FAD is now lower than that of UNAGAN. Together with the result in JS and NDB, we see from this table that StyleGAN2 is more effective in learning to cover the modes in a large dataset.
Figure \ref{fig:melspec} demonstrates the mel-spectrograms of some random drum loops generated by this StyleGAN2 model.


\subsection{Subjective Evaluation \& Its Result}

We run additionally an online listening test to evaluate the models subjectively.
Each subject is presented with the a randomly-picked human-made loop from the freesound dataset, and one randomly-generated loop by each of the three models trained on \emph{freesound}, with the ordering of these four loops randomized.
Then, the subject is asked to rate each of these one-bar loops in terms of the following metrics, the first three on a three-point scale, and the last one on a five-point Likert scale:
\begin{itemize}[leftmargin=*,itemsep=0pt,topsep=2pt]
    \item \textbf{Drumness}:
    whether the sample contains drum sounds (`no'/`yes but vague'/`yes and clear');
    \item \textbf{Loopness}:
    whether the sample can be played repeatedly in a seamless manner (`no'/`yes but not so good'/`yes');
    \item \textbf{Audio quality}: whether the sample is free of unpleasant noises or artifacts (`no'/`no but not so bad'/`yes');
    \item \textbf{Preference}: how much you like it (1--5).
\end{itemize}
To evaluate loopness, we actually repeat each sample four times in the audio recording presented to the subjects. 
And, since the output of the models go through the MelGAN vocoder to become waveforms, we compute the Mel spectrograms of the human-made loops and render them to audio with the same vocoder for fair comparison.

\begin{table}[t]
\centering
\begin{tabular}{ |l | r r r r |} 
 \hline
& \textbf{IS}~$\uparrow$  & \textbf{FAD}~$\downarrow$ & \textbf{JS}~$\downarrow$ & \textbf{NDB/$K$}~$\downarrow$\\ 
 \hline
 Looperman &11.9\footnotesize{$\pm$3.21}  & 0.11  & 0.01 & 0.01 \\ 
 Freesound & 6.30\footnotesize{$\pm$1.82}  & 0.72  & 0.08 & 0.46 \\ \hline
 StyleGAN & 1.31\footnotesize{$\pm$1.95}  & 13.78 & 0.43 & 0.94 \\ 
 StyleGAN2 & \textbf{5.24\footnotesize{$\pm$1.84}}  & 7.91  & \textbf{0.09} & \textbf{0.59}   \\
 UNAGAN  & 3.33\footnotesize{$\pm$1.65}  & \textbf{4.32}  & 0.16  & 0.73 \\
 \hline
\end{tabular}
\caption{Objective evaluation result for the three models trained on the \emph{freesound} dataset. We also display the IS of the two sets of real data. ($\downarrow$\,/\,$\uparrow$: the lower/higher the better).}
\label{tab:exp-obj-freesound}
\end{table}
\begin{table}[t]
\centering
\begin{tabular}{ |l | r r r r |} 
 \hline
 & \textbf{IS}~$\uparrow$  & \textbf{FAD}~$\downarrow$ & \textbf{JS}~$\downarrow$ & \textbf{NDB/$K$}~$\downarrow$\\ 
 \hline
 StyleGAN  & 1.30\footnotesize{$\pm$2.00}  & 12.98 & 0.41 & 0.87\\ 
 StyleGAN2 & \textbf{6.08\footnotesize{$\pm$2.26}}  & \textbf{2.22}  & \textbf{0.01} & \textbf{0.08} \\
 UNAGAN    & 3.83\footnotesize{$\pm$1.72}  & 3.36  & 0.29 & 0.89 \\
 \hline
\end{tabular}
\caption{Objective evaluation result for the three models trained on instead the private \emph{looperman} dataset.}
\label{tab:exp-obj-looperman}
\end{table}

\begin{figure}[t]
\hspace*{-0.3cm}
\vspace{-0.3cm}
\includegraphics[clip,width=1\columnwidth]{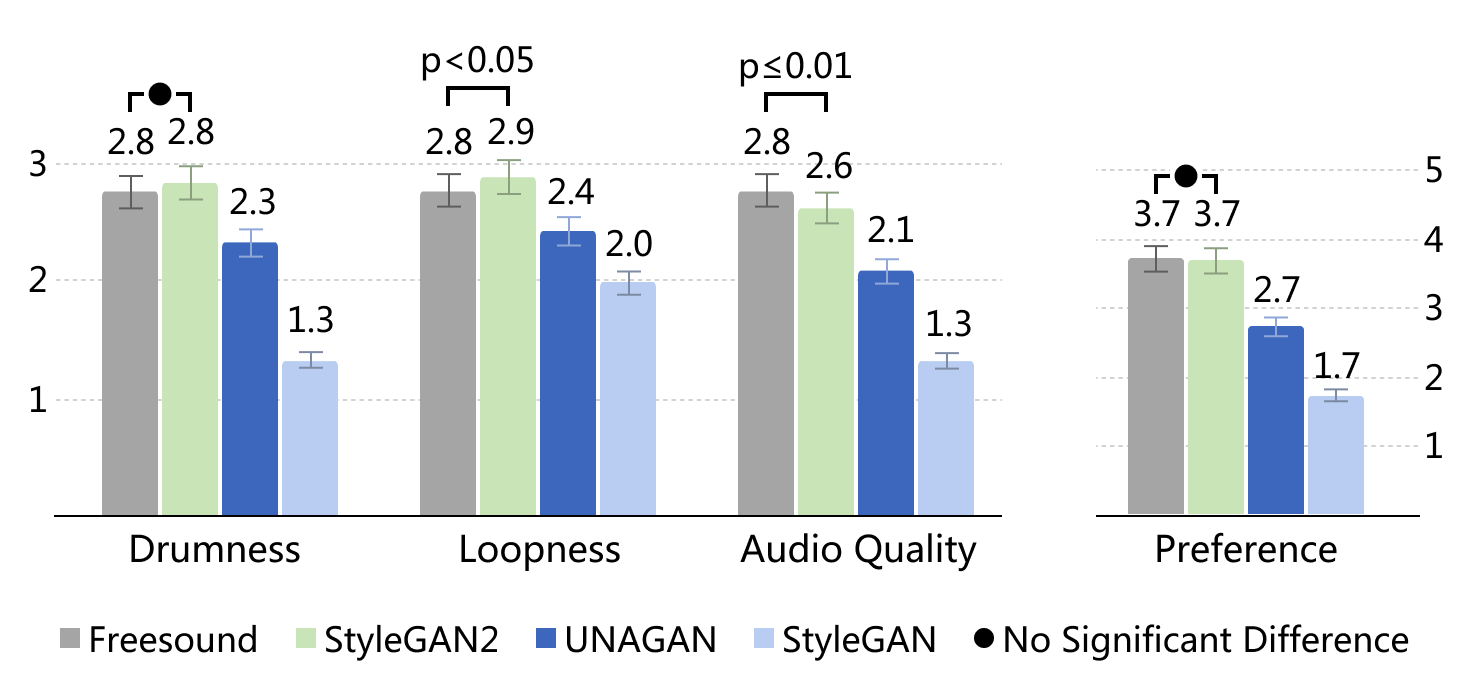}
\caption{Subjective evaluation result for the three models trained on \emph{freesound}. The performance difference between any pair of models in any metric is statistically significant ($p$-value$<0.001$) under the Wilcoxon signed-rank test,
except for the pairs that are explicitly highlighted.} 
\label{fig:subjective}
\end{figure}

140 anonymous subjects from Taiwan participated in this test,\footnote{The subjects have no ideas about our models beforehand; they neither know that one of the loops they hear is human-made.} 
with in total six unique samples by each model evaluated.
Overall, the responses indicated an acceptable level of reliability (Cronbach’s $\alpha=0.709$).
We see from Figure \ref{fig:subjective} 
that the result of this subjective evaluation is well aligned with that of the objective evaluation, with StyleGAN2 performing the best and StyleGAN the worst, demonstrating the effectiveness of the objective metrics to some extent.
Interestingly, we see no statistical difference in the ratings of the StyleGAN2 loops and the (MelGAN-vocoded) freesound loops in Drumness and Preference.


\yhyangtext{Finally, we correlate the scores of the objective metrics and subjective metrics for the 18 samples evaluated in the listening test (i.e., six samples by each GAN model). We found 0.25--0.37 correlation between IS and the four subjective metrics, and 0.01--0.16 negative correlation between FAD and the subjective metrics. The strongest correlation (0.37) is found between IS and Preference.}

\section{Conclusion and Future Work}
\label{sec:conclusion}

In this paper, we have proposed using loop generation as a benchmarking task to provide a standardized evaluation of audio-domain music generation models, taking advantage of the public availability of the large collection of loops in FSLD.
Moreover, we developed customized metrics to objectively evaluate the performance of such generative models for the particular case of one-bar drum loops with 120 BPM.
As references, we implemented and evaluated three recent model architectures using the dataset, and discovered that StyleGAN2 works quite well.
The list of models we have evaluated is short and by no means exhaustive. We wish researchers can find this benchmark useful and consider it as part of their evaluation of new models.

This work can be extended in many other directions. First and foremost, we can extend the benchmark to cover all the loops in FSLD (and looperman). The major complexity here could be the challenge to build a model that fits it all; we may need separate generative models and vocoders for different types of loops.

Second, we are certainly interested in the case of generating loops that have different tempos, rather than a fixed tempo at 120. This will require the generative models to be capable of generating variable-length output, which seems more realistic in musical audio applications. 

We can also extend the benchmark to generate four-bar loops (which are not simply repeating a one-bar loop quadruple times), as there are actually a big collection of 6,656 four-bar drum loops in the looperman dataset. We do not evaluate this in this paper, as the public freesound dataset does not contain many such four-bar loops.

We also want to include more objective metrics in the future, such as using the Audio Commons Audio Extractor \cite{audiocommons} to evaluate the ``loopness'' of the generated samples, or using 
an automatic drum transcription model \cite{8350302,choi20ismir,callender20arxiv} to assess the plausibility of the created percussive patterns.

Besides the benchmarking initiative, we are interested in further improving  audio-domain loop generation itself and exploring new use cases, e.g., to have a conditional generation model that gives users some control (in similar veins to \cite{drumgan,9053128, loopnet}), or to aim at generating novel loops by means of a creative adversarial network (CAN) \cite{tokui2020gan}. 

\section{Acknowledgements}
This research work is supported by the Ministry of Science and Technology (MOST), Taiwan, under grant number 109-2628-E-001-002-MY2.

\bibliography{hung21ismir_CR}

\begin{thebibliography}{10}
\providecommand{\url}[1]{#1}
\csname url@samestyle\endcsname
\providecommand{\newblock}{\relax}
\providecommand{\bibinfo}[2]{#2}
\providecommand{\BIBentrySTDinterwordspacing}{\spaceskip=0pt\relax}
\providecommand{\BIBentryALTinterwordstretchfactor}{4}
\providecommand{\BIBentryALTinterwordspacing}{\spaceskip=\fontdimen2\font plus
\BIBentryALTinterwordstretchfactor\fontdimen3\font minus
  \fontdimen4\font\relax}
\providecommand{\BIBforeignlanguage}[2]{{%
\expandafter\ifx\csname l@#1\endcsname\relax
\typeout{** WARNING: IEEEtran.bst: No hyphenation pattern has been}%
\typeout{** loaded for the language `#1'. Using the pattern for}%
\typeout{** the default language instead.}%
\else
\language=\csname l@#1\endcsname
\fi
#2}}
\providecommand{\BIBdecl}{\relax}
\BIBdecl

\bibitem{engel2017neural}
J.~Engel \emph{et~al.}, ``Neural audio synthesis of musical notes with
  {WaveNet} autoencoders,'' in \emph{Proc. Int. Conf. Machine Learning}, 2017.

\bibitem{engel2019gansynth}
J.~Engel, K.~K. Agrawal, S.~Chen, I.~Gulrajani, C.~Donahue, and A.~Roberts,
  ``{GANS}ynth: Adversarial neural audio synthesis,'' in \emph{Proc. Int. Conf.
  Learning Representations}, 2019.

\bibitem{drumgan}
J.~Nistal, S.~Lattner, and G.~Richard, ``{DrumGAN}: Synthesis of drum sounds
  with timbral feature conditioning using generative adversarial networks,'' in
  \emph{Proc. Int. Soc. Music Information Retrieval Conf.}, 2020.

\bibitem{9053128}
A.~Ramires, P.~Chandna, X.~Favory, E.~Gómez, and X.~Serra, ``Neural percussive
  synthesis parameterised by high-level timbral features,'' in \emph{Proc. IEEE
  Int. Conf. Acoustics, Speech and Signal Processing}, 2020.

\bibitem{oord16}
A.~van~den Oord \emph{et~al.}, ``{WaveNet}: A generative model for raw audio,''
  \emph{arXiv preprint arXiv:1609.03499}, 2016.

\bibitem{samplernn}
S.~Mehri \emph{et~al.}, ``{SampleRNN}: An unconditional end-to-end neural audio
  generation model,'' in \emph{Proc. Int. Conf. Learning Representations},
  2017.

\bibitem{dadabots}
C.~J. Carr and Z.~Zukowski, ``Generating albums with {SampleRNN} to imitate
  metal, rock, and punk bands,'' \emph{arXiv preprint:1811.06633}, 2018.

\bibitem{melnet}
S.~Vasquez and M.~Lewis, ``{MelNet}: A generative model for audio in the
  frequency domain,'' \emph{arXiv preprint arXiv:1906.01083}, 2019.

\bibitem{liu2020unconditional}
J.-Y. Liu, Y.-H. Chen, Y.-C. Yeh, and Y.-H. Yang, ``Unconditional audio
  generation with generative adversarial networks and cycle regularization,''
  in \emph{Proc. INTERSPEECH}, 2020.

\bibitem{liu2019score}
------, ``Score and lyrics-free singing voice generation,'' in \emph{Proc. Int.
  Conf. Computational Creativity}, 2020.

\bibitem{dhariwal2020jukebox}
P.~Dhariwal, H.~Jun, C.~Payne, J.~W. Kim, A.~Radford, and I.~Sutskever,
  ``Jukebox: A generative model for music,'' \emph{arXiv preprint:2005.00341},
  2020.

\bibitem{mp3net}
K.~van~den Broek, ``{MP3net}: coherent, minute-long music generation from raw
  audio with a simple convolutional {GAN},'' \emph{arXiv preprint
  arXiv:2101.04785}, 2021.

\bibitem{9156570}
T.~Karras \emph{et~al.}, ``Analyzing and improving the image quality of
  {StyleGAN},'' in \emph{Proc. IEEE/CVF Conf. Computer Vision and Pattern
  Recognition}, 2020.

\bibitem{won2020evaluation}
M.~Won, A.~Ferraro, D.~Bogdanov, and X.~Serra, ``Evaluation of {CNN}-based
  automatic music tagging models,'' in \emph{Proc. Sound and Music Computing
  Conf.}, 2020.

\bibitem{ramires2020freesound}
A.~Ramires \emph{et~al.}, ``{The Freesound Loop Dataset} and annotation tool,''
  in \emph{Proc. Int. Soc. Music Information Retrieval Conf.}, 2020.

\bibitem{LoopsasGenreResources}
\BIBentryALTinterwordspacing
G.~Stillar, ``Loops as genre resources,'' \emph{Folia Linguistica}, vol.~39,
  no. 1-2, pp. 197 -- 212, 2005. [Online]. Available:
  \url{https://www.degruyter.com/view/journals/flin/39/1-2/article-p197.xml}
\BIBentrySTDinterwordspacing

\bibitem{joannching}
J.~Ching, A.~Ramires, and Y.-H. Yang, ``Instrument role classification:
  Auto-tagging for loop based music,'' in \emph{Proc. Joint Conference on AI
  Music Creativity}, 2020.

\bibitem{GAN}
I.~J. Goodfellow \emph{et~al.}, ``Generative adversarial nets,'' in \emph{Proc.
  Advances in Neural Information Processing Systems}, 2014, pp. 2672--2680.

\bibitem{8953766}
T.~Karras, S.~Laine, and T.~Aila, ``A style-based generator architecture for
  generative adversarial networks,'' in \emph{Proc. IEEE/CVF Conf. Computer
  Vision and Pattern Recognition}, 2019, pp. 4396--4405.

\bibitem{paul20ismir}
B.-Y. Chen, J.~Smith, and Y.-H. Yang, ``Neural loop combiner: Neural network
  models for assessing the compatibility of loops,'' in \emph{Proc. Int. Soc.
  Music Information Retrieval Conf.}, 2020.

\bibitem{richard17}
R.~Vogl and P.~Knees, ``An intelligent drum machine for {Electronic Dance
  Music} production and performance,'' in \emph{Proc. Int. Conf. New Interfaces
  for Musical Expression}, 2017.

\bibitem{DBLP:conf/nips/SalimansGZCRCC16}
T.~Salimans \emph{et~al.}, ``Improved techniques for training {GANs},'' in
  \emph{Proc. Conf. Neural Information Processing Systems}, 2016, pp.
  2226--2234.

\bibitem{nistal2021vqcpcgan}
J.~Nistal, C.~Aouameur, S.~Lattner, and G.~Richard, ``{VQCPC-GAN}:
  Variable-length adversarial audio synthesis using vector-quantized
  contrastive predictive coding,'' in \emph{Proc. IEEE Workshop on Applications
  of Signal Processing to Audio and Acoustics}, 2021.

\bibitem{aouameur2019neural}
C.~Aouameur, P.~Esling, and G.~Hadjeres, ``Neural drum machine: An interactive
  system for real-time synthesis of drum sounds,'' \emph{arXiv preprint
  arXiv:1907.02637}, 2019.

\bibitem{drysdale2020adversarial}
J.~Drysdale, M.~Tomczak, and J.~Hockman, ``Adversarial synthesis of drum
  sounds,'' in \emph{Proc. Int. Conf. Digital Audio Effects}, 2020.

\bibitem{loopnet}
P.~Chandna, A.~Ramires, X.~Serra, and E.~Gómez, ``{LoopNet}: Musical loop
  synthesis conditioned on intuitive musical parameters,'' in \emph{Proc. IEEE
  Int. Conf. Acoustics, Speech and Signal Processing}, 2021.

\bibitem{DBLP:conf/icml/GillickREEB19}
J.~Gillick, A.~Roberts, J.~Engel, D.~Eck, and D.~Bamman, ``Learning to groove
  with inverse sequence transformations,'' in \emph{Proc. Int. Conf. Machine
  Learning}, 2019.

\bibitem{drumVAE}
V.~Thio, H.-M. Liu, Y.-C. Yeh, and Y.-H. Yang, ``A minimal template for
  interactive web-based demonstrations of musical machine learning,'' in
  \emph{Proc. Workshop on Intelligent Music Interfaces for Listening and
  Creation}, 2019.

\bibitem{alain2020deepdrummer}
G.~Alain, M.~Chevalier-Boisvert, F.~Osterrath, and R.~Piche-Taillefer,
  ``{DeepDrummer}: Generating drum loops using deep learning and a human in the
  loop,'' \emph{arXiv preprint arXiv:2008.04391}, 2020.

\bibitem{tokui2020gan}
N.~Tokui, ``Can {GAN} originate new electronic dance music genres? --
  {Generating} novel rhythm patterns using {GAN} with genre ambiguity loss,''
  \emph{arXiv preprin: 2011.13062}, 2020.

\bibitem{musegan}
H.-W. Dong, W.-Y. Hsiao, L.-C. Yang, and Y.-H. Yang, ``{MuseGAN}:
  Symbolic-domain music generation and accompaniment with multi-track
  sequential generative adversarial networks,'' in \emph{Proc. AAAI Conf.
  Artificial Intelligence}, 2018.

\bibitem{simon2018learning}
I.~Simon, A.~Roberts, C.~Raffel, J.~Engel, C.~Hawthorne, and D.~Eck, ``Learning
  a latent space of multitrack measures,'' \emph{arXiv preprint
  arXiv:1806.00195}, 2018.

\bibitem{ren2020popmag}
Y.~Ren, J.~He, X.~Tan, T.~Qin, Z.~Zhao, and T.-Y. Liu, ``Pop{MAG}: Pop music
  accompaniment generation,'' in \emph{Proc. ACM Multimedia Conf.}, 2020.

\bibitem{audiocommons}
F.~Font, T.~Brookes, G.~Fazekas, M.~Guerber, A.~La~Burthe, D.~Plans,
  M.~Plumbley, W.~Wang, and X.~Serra, ``{Audio Commons}: {Bringing} {Creative
  Commons} audio content to the creative industries,'' in \emph{Proc. AES Int.
  Conf. Audio for Games,}, 2016.

\bibitem{lattner2019highlevel}
S.~Lattner and M.~Grachten, ``High-level control of drum track generation using
  learned patterns of rhythmic interaction,'' in \emph{Proc. IEEE Work.
  Applications of Signal Processing to Audio and Acoustics}, 2019.

\bibitem{barratt2018note}
S.~Barratt and R.~Sharma, ``A note on the inception score,'' in \emph{Proc.
  ICML Works. Theoretical Foundations and Applications of Deep Generative
  Models}, 2018.

\bibitem{10.1145/2964284.2973795}
S.~B\"{o}ck, F.~Korzeniowski, J.~Schl\"{u}ter, F.~Krebs, and G.~Widmer,
  ``Madmom: A new {Python} audio and music signal processing library,'' in
  \emph{Proc. ACM Multimedia Conf.}, 2016.

\bibitem{Bock2016d}
S.~B{\"{o}}ck, F.~Krebs, and G.~Widmer, ``Joint beat and downbeat tracking with
  recurrent neural networks,'' in \emph{Proc. Int. Soc. Music Information
  Retrieval Conf.}, 2016.

\bibitem{ballroom}
F.~Gouyon, A.~Klapuri, S.~Dixon, M.~Alonso, G.~Tzanetakis, C.~Uhle, and
  P.~Cano, ``An experimental comparison of audio tempo induction algorithms,''
  \emph{IEEE Trans. Audio, Speech, and Language Processing}, vol.~14, no.~5,
  pp. 1832--1844, 2006.

\bibitem{kilgour2019frechet}
K.~Kilgour, M.~Zuluaga, D.~Roblek, and M.~Sharifi, ``{Fr\'echet Audio
  Distance}: A metric for evaluating music enhancement algorithms,''
  \emph{arXiv preprint arXiv: 1812.08466}, 2019.

\bibitem{DBLP:conf/nips/RichardsonW18}
E.~Richardson and Y.~Weiss, ``On {GANs} and {GMMs},'' in \emph{Proc. Conf.
  Neural Information Processing Systems}, 2018.

\bibitem{karras2018progressive}
T.~Karras, T.~Aila, S.~Laine, and J.~Lehtinen, ``Progressive growing of {GAN}s
  for improved quality, stability, and variation,'' in \emph{Proc. Int. Conf.
  Learning Representations}, 2018.

\bibitem{palkama20interspeech}
K.~Palkama, L.~Juvela, and A.~Ilin, ``Conditional spoken digit generation with
  {StyleGAN},'' in \emph{Proc. INTERSPEECH}, 2020.

\bibitem{prenger2018waveglow}
R.~Prenger, R.~Valle, and B.~Catanzaro, ``{WaveGlow}: A flow-based generative
  network for speech synthesis,'' in \emph{Proc. IEEE Int. Conf. Acoustics,
  Speech and Signal Processing}, 2019.

\bibitem{kong2021diffwave}
Z.~Kong, W.~Ping, J.~Huang, K.~Zhao, and B.~Catanzaro, ``{DiffWave}: A
  versatile diffusion model for audio synthesis,'' in \emph{Proc. Int. Conf.
  Learning Representations}, 2021.

\bibitem{melgan}
K.~Kumar, R.~Kumar, T.~de~Boissiere, L.~Gestin, W.~Z. Teoh, J.~Sotelo,
  A.~de~Brebisson, Y.~Bengio, and A.~Courville, ``{MelGAN}: Generative
  adversarial networks for conditional waveform synthesis,'' \emph{arXiv
  preprint arXiv:1910.06711}, 2019.

\bibitem{mescheder2018training}
L.~Mescheder, A.~Geiger, and S.~Nowozin, ``Which training methods for gans do
  actually converge?'' \emph{arXiv preprint arXiv:1801.04406}, 2018.

\bibitem{donahue19iclr}
C.~Donahue, J.~McAuley, and M.~Puckette, ``{Adversarial audio synthesis},'' in
  \emph{Proc. Int. Conf. Learning Representations}, 2019.

\bibitem{berthelot17}
D.~Berthelot, T.~Schumm, and L.~Metz, ``{BEGAN}: Boundary equilibrium
  generative adversarial networks,'' \emph{arXiv preprint:1703.10717}, 2017.

\bibitem{cyclegan}
J.-Y. Zhu, T.~Park, P.~Isola, and A.~A. Efros, ``Unpaired image-to-image
  translation using cycle-consistent adversarial networks,'' in
  \emph{Proceedings of the IEEE international conference on computer vision},
  2017.

\bibitem{8350302}
C.-W. Wu, C.~Dittmar, C.~Southall, R.~Vogl, G.~Widmer, J.~Hockman, M.~Müller,
  and A.~Lerch, ``A review of automatic drum transcription,'' \emph{IEEE/ACM
  Trans. Audio, Speech, and Language Processing}, vol.~26, no.~9, pp.
  1457--1483, 2018.

\bibitem{choi20ismir}
K.~Choi and K.~Cho, ``Deep unsupervised drum transcription,'' in \emph{Proc.
  Int. Soc. Music Information Retrieval Conf.}, 2020.

\bibitem{callender20arxiv}
L.~Callender, C.~Hawthorne, and J.~H. Engel, ``Improving perceptual quality of
  drum transcription with the expanded groove {MIDI} dataset,'' \emph{arXiv
  preprint:2004.00188}, 2020.

\end{thebibliography}

\end{document}